# Human Dynamics:
# The Correspondence Patterns of Darwin and Einstein

While living in different historical era, Charles Darwin (1809–1882) and Albert Einstein (1879–1955) were both prolific correspondents: Darwin sent (received) at least 7,591 (6,530) letters during his lifetime while Einstein sent (received) over 14,500 (16,200). Before email scientists were part of an extensive university of letters, the main venue for exchanging new ideas and results. But were the communication patterns of the pre-email times any different from the current era of instant access? Here we show that while the means have changed, the communication dynamics has not: Darwin's and Einstein's pattern of correspondence and today's electronic exchanges follow the same scaling laws. Their communication belongs, however, to a different universality class from email communication, providing evidence for a new class of phenomena capturing human dynamics.

We start from a record containing the sender, recipient and the date of each letter[1,2] sent or received by the two scientists. As Fig 1a indicates, their correspondence exploded after their raise to fame, and reached a highly fluctuating yet relatively steady pattern afterwards. While on average they write 0.59 (D) and 1.02 (E) letters per day during the last 30 years of their life, these averages hide significant daily fluctuations. For example Darwin writes 12 letters on 1874-1-1 and Einstein receives 120 letters on 1949-3-14, for his 70[th] birthday.

The response time, $\tau$, represents the time interval between the date of a letter received from a given person, and the date of the next letter from Darwin or Einstein to him or her. As shown in Fig. 1b and c, the probability that a letter will be replied to in $\tau$ days is well approximated by a power law $P(\tau) \sim \tau^{-\alpha}$ with $\alpha = 3/2$. The fact that the scaling spans close to four orders of magnitude, from days to years, indicates that the majority of responses (53%E, 63%D) took less than ten days. In some cases, however, the correspondence was stalled for months or years. Some of these represent long breaks in the correspondence and a few are a consequence of missing letters. Others, however, correspond to genuine delays, like Einstein's Oct 14, 1921 response to Ralph De Laer Kronig's letter of Sept. 1920, starting with "in the course of eating myself through a mountain of correspondence I find your interesting letter from September of last year."

To understand the origin of the observed scaling behavior we have to realize that given the wide range of response times, both Darwin and Einstein must have prioritized their correspondence, responding to the received letters based on priorities they assigned to each. Thus a simple model of their correspondence assumes that letters arrive at a rate $\lambda$ and are responded to at a rate $\mu$. Each letter is assigned a priority, always the highest priority letter being answered next. Therefore, high priority letters are responded soon after their arrival, while low priority letters will have to wait considerable time intervals. The waiting time distribution of this simple model[3] follows[4] $P(\tau) \sim \tau^{-3/2} \exp(-\tau/\tau_0)$, which predicts a power law waiting time for the critical regime $\lambda = \mu$, when $\tau_0 = \infty$. Given that Einstein and Darwin answer only a fraction of letters they receive (their overall response rate being 0.32 (D) and 0.24 (E)), we have $\lambda > \mu$, placing the model in the supercritical regime, where a finite fraction of letters are never answered. Numerical simulations indicate that in this supercritical regime the waiting time distribution of the responded letters also follows a power law with exponent $\alpha = 3/2$, different from $\alpha = 1$ obtained for email communications[5]. Therefore, while the response times in email and mail communications follow the same scaling law, they belong to different universality classes.

The correspondence pattern of Einstein and Darwin, beyond representing examples of well-mapped human interaction patterns, is of major historical interest as well. The fact that they did answer most letters in a timely fashion, indicates that while they had many other responsibilities, they were acutely aware of the importance of this intellectual intercourse. Yet, the occasional delays prioritizing forced on them were not always without consequences. For example, on Oct 14, 1921 Einstein returns to a correspondence with Theodor Kaluza that he left off two years earlier, when he discouraged Kaluza from publishing one his papers. Having second thoughts, he recommends the paper's submission. Encouraged by Einstein's two years late change of mind, Kaluza does so, submitting the now famous paper on five-dimensional unified field theory[6], a key component of today's string theory. Would it have made a difference for the course of science if Einstein does not waver for two years about it? We will never know. But our results indicate that Darwin's and Einstein's late responses or resumed correspondences are not singularities or exceptions: they are part of a universal scaling law[7], representing a fundamental pattern of human dynamics that neither the famous, nor the undistinguished can escape.

**João Gama Oliveira\* +, Albert-László Barabási\* ‡**
\*Center for Complex Network Research and Department of Physics, University of Notre Dame, IN 46556 USA
+Departamento de Física, Universidade de Aveiro, 3810-193 Aveiro, Portugal
‡ Center for Cancer Systems Biology, Dana-Farber Cancer Institute, Harvard University, Boston, MA 02115
e-mail: alb@nd.edu

**Supplementary Information** accompanies the communication on www.nature.com/nature.
**Competing financial interests:** declared none.

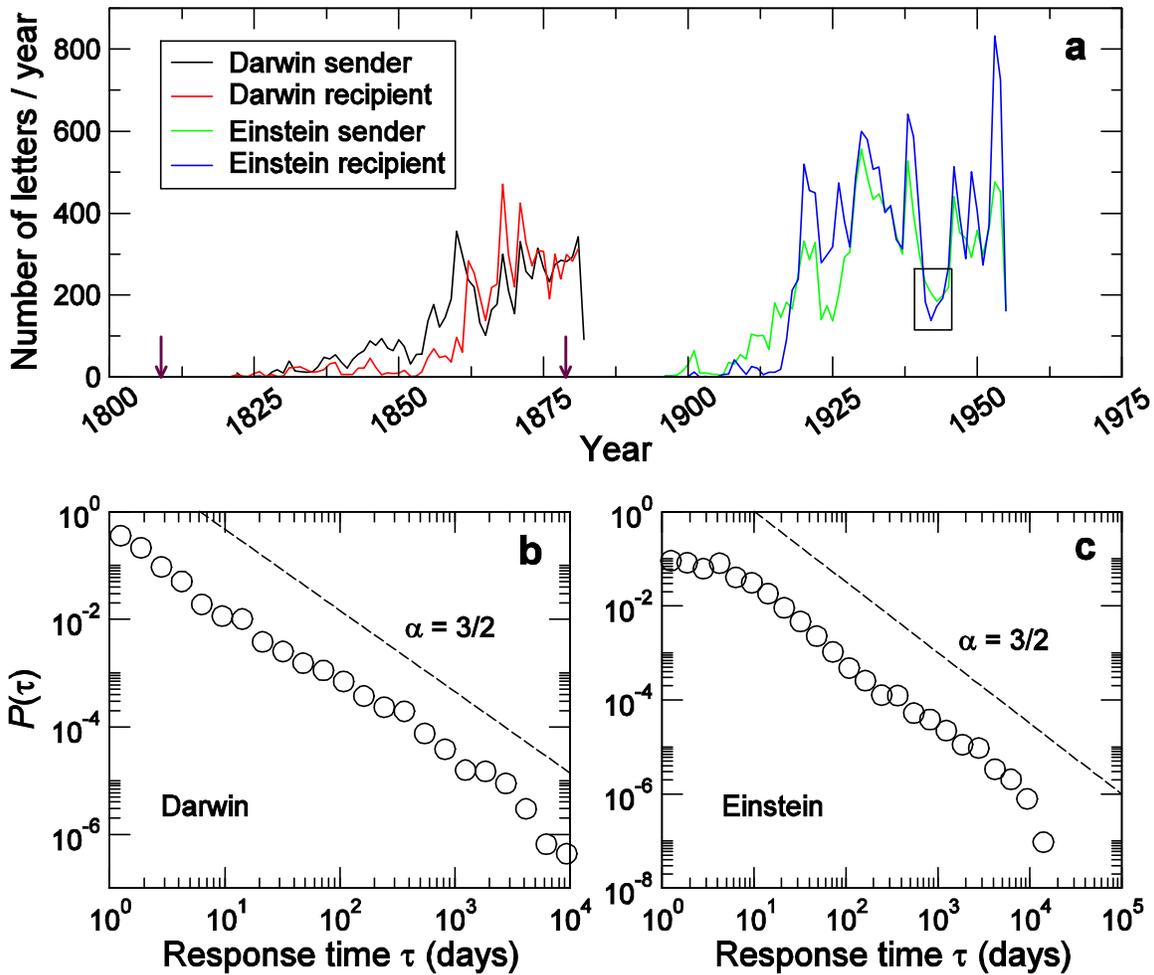

**Figure 1** The correspondence patterns of Darwin and Einstein. **a.** The historical record of the number of letters sent/received each year based on their full correspondence[1,2]. An anomalous drop in Einstein's correspondence marks the Second World War period (1939–1945, see box). The arrows mark the birth dates of Darwin and Einstein. **b** and **c.** Distribution of the response times for the letters replied to by Darwin and Einstein, respectively. Note that both distributions are well approximated with a power law tail with exponent $\alpha = 3/2$, the best fit over the whole data for Darwin providing $\alpha = 1.45 \pm 0.1$ and for Einstein $\alpha = 1.47 \pm 0.1$.